\documentclass[12pt]{article}
\usepackage{lineno,hyperref}
\usepackage{geometry}
\usepackage{amssymb, graphics, setspace}
\usepackage{pgfplots}
\usepackage{textcomp}
\usepackage[caption=false]{subfig}
\usepackage{indentfirst}
\usepackage{amsmath}
\usepackage{commath}
\usepackage{etoolbox}
\makeatletter
\usepackage{graphicx,bm}
\usepackage{url}
\usepackage{float}
\usepackage{textcomp}
\usepackage{bbold}
\usepackage{verbatim}

\usepackage{changes}
\usepackage{textcomp}
\usepackage{authblk}

\usepackage{fourier}
\begin{document}
\title{Regge models of proton diffractive dissociation based on factorization and structure functions}

\author[1]{L\'aszl\'o Jenkovszky \thanks{jenk@bitp.kiev.ua}}
\author[2]{Rainer Schicker \thanks{schicker@physi.uni-heidelberg.de}}
\author[3]{Istv\'an Szanyi \thanks{szanyi.istvan@wigner.hu}}

\affil[1]{\small  Bogolyubov Institute for Theoretical Physics (BITP), \protect\\ 
Ukrainian National Academy of Sciences, \protect\\  14-b, Metrologicheskaya str.,
Kiev, 03680, Ukraine}
\affil[2]{Physikalisches Institut, University Heidelberg, \protect\\ Im Neuenheimer Feld 226,
69120 Heidelberg, Germany}
\affil[3]{E\"otv\"os University, H-1117 Budapest, P\'azm\'any P. s. 1/A, Hungary;\protect\\ Wigner Research Centre for Physics, H-1525 Budapest 114, POBox 49, Hungary;\protect\\ MATE Institute of Technology,  K\'aroly R\'obert Campus, H-3200 Gy\"ongy\"os, M\'atrai \'ut 36, Hungary}

\maketitle

\abstract{Recent results by the authors on proton diffractive dissociation (single, double and central) in the low-mass resonance region with emphasis on the LHC kinematics are reviewed and updated. Based on the previous ideas that the contribution of the inelastic proton-Pomeron vertex can be described by the proton structure function, the contribution of the inelastic Pomeron-Pomeron vertex appearing in central diffraction is now described by a Pomeron structure function.}

\section{Introduction}
Proton and deuteron diffractive dissociation was intensively studied in the past century at FNAL and at CERN-ISR. The relevant experimental results and their phenomenological interpretation were covered in a large number of papers, see Refs. 
\cite{Goulianos:1982vk, Goulianos:2001hck, Goulianos:2004as} and references therein. Recent LHC-related developments are discussed e.g. in Refs.\cite{Goulianos:1995vn, Goulianos:2018atd,Ciesielski:2012mc}. 
The basic idea behind these and similar studies is the identification of the exchanged Pomeron with  a flux emitted by the diffractively scattered proton \cite{Ingelman:1984ns}.

A different  point of view was taken in Refs. \cite{Jenkovszky:2010ym, Jenkovszky:2011bt, Jenkovszky:2012hf, Jenkovszky:2013xny}, where, following C.A. Jaroszkiewicz and P.V. Landshoff \cite{PhysRevD.10.170}, the unknown inelastic proton-Pomeron ($pP$) vertex was associated with the deep inelastic scattering (DIS) photon-nucleon structure function (SF), known from the experiments at HERA. In doing so, G.A. Jaroszkiewicz and P.V. Landshoff \cite{PhysRevD.10.170} used a high-energy, Regge-behaved formula for the DIS SF, leaving outside the low-energy (missing mass) resonance structure. Resonances were included in this formalism in a series of papers \cite{Jenkovszky:2010ym, Jenkovszky:2011bt, Jenkovszky:2012hf, Jenkovszky:2013xny}, where by duality the high-energy behaviour of the SF was replaced by its low-energy (missing mass) SF, dominated by direct channel non-linear complex Regge trajectories, producing finite-widths resonances. Now we extend the structure function formalism to inelastic Pomeron-Pomeron ($PP$) vertex to model central diffractive processes.

Diffractive dissociation is interesting and important for many reasons. One is that new experimental data are expected  from ongoing the LHC run, especially in the  central region, that will help us to fix the remaining freedom/flexibility of the models. On the other hand, the predictions of the model may guide experimentalists in tuning their detectors. Also, it is important to remember that the high energies, typical of the LHC make possible to neglect in most of the kinematical configurations the contribution from secondary reggeons and allow us to use Regge factorisation and concentrate on the nature of the Pomeron.

The paper is organized as follows: in Sec. \ref{sec:models} models of differential cross sections of the diffractive processes, including elastic scattering as well as single, double and central diffractive dissociation are constructed. In Sec. \ref{sec:proton_sf} the treatment of the $pP$ and $PP$ vertices is introduced based on the formalism of structure functions. The calculated integrated cross sections for processes with diffractive dissociation including fits to the available measured data are presented in Sec. \ref{sec:intcs}. The calculated differential cross sections are presented in Sec. \ref{sec:diffcs}. Our results and the conclusions are summarised in Sec.~\ref{sec:concl}.

\section{Differential cross sections}\label{sec:models}
In this section we summarize and update the basic formulae for elastic scattering, single diffractive dissociation and double diffractive dissociation (elaborated in a series of papers \cite{Szanyi:2019kkn,Jenkovszky:2010ym, Jenkovszky:2011bt, Jenkovszky:2012hf, Jenkovszky:2013xny}), also we extend the formalism based on the use of  structure functions to central diffractive dissociation and mixed processes. This is an important step on the way towards the elaboration of a unique and adequate language and relevant set of variables and measurables, understandable and convenient both for theorists and experimentalists.

\begin{figure}[h]
\begin{center}
\includegraphics[clip,scale=0.7,]{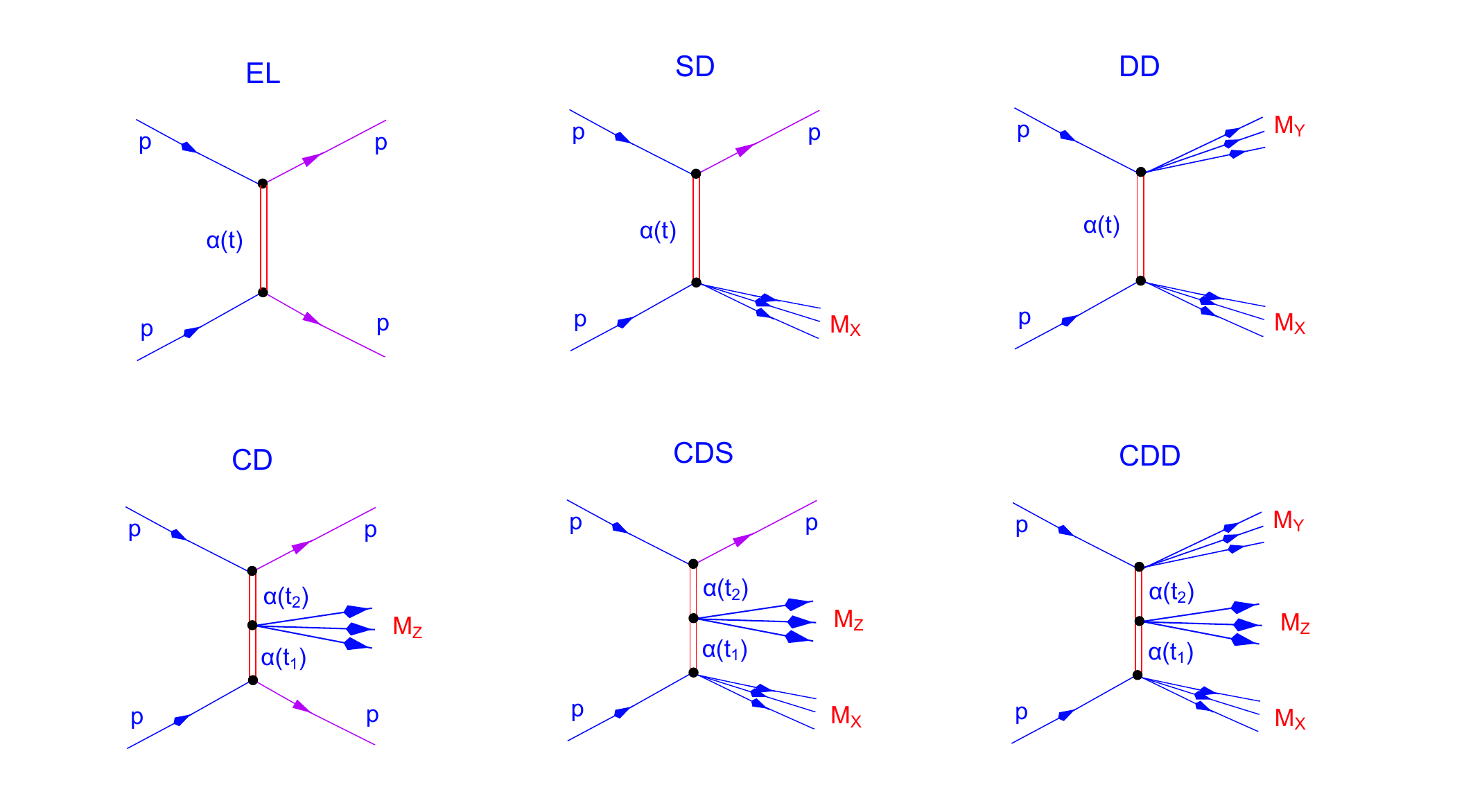}
\end{center}
%\vspace{-10cm}
\caption{Diffraction: elastic ($EL$); single ($SD$), double $(DD)$ and central ($CD$) dissociation; mixed central and single dissociation ($CDS$);  mixed central and double dissociation ($CDD$).}
\label{Fig:DD}
\end{figure}

Fig. \ref{Fig:DD} shows the main topologies appearing in diffractive dissociation under discussion. It may serve also as a guide to relevant equations that follow.   

The differential cross section of elastic (EL) proton-proton scattering is:
\begin{equation}
\frac{d\sigma_{EL}}{dt}=A_{EL}\beta^2(t)\beta^2(t)|\eta(t)|^2\left(\frac{s}{s_0}\right)^{2\alpha_{P}(t)-2},
\end{equation}
where $A_{EL}$ is a free parameter\footnote{$A_i$ with $i=EL$ and, later in the text, $i\in\{SD,~DD,~CD,~CDS,~CDD\}$ are free parameters of dimension $\left[A_{i}\right]$ = mb/GeV$^2$, including also normalisation constants.}, $s$ and $t$ are the Mandelstam variables. The proton-Pomeron coupling squared is: \mbox{$\beta^2(t)=e^{bt}$}, where $b$ is a free parameter and $b\approx1.97$  GeV$^{-2}$ determined in Ref. \cite{Szanyi:2019kkn}. The Pomeron trajectory is $\alpha_{P}(t)=1+\epsilon+\alpha't$, where $\epsilon\approx0.08$ and $\alpha'\approx0.3$ GeV$^{-2}$ \cite{Szanyi:2019kkn}. The signature factor is $\eta(t)=e^{-i\frac{\pi}{2}\alpha_P(t)}$; its contribution to the differential cross section is  $|\eta(t)|^2=1$, therefore we ignore it in what follows. We set $s_0=1$ GeV$^2$ for simplicity.

The differential cross section of  proton-proton single diffraction (SD) is:
\begin{equation}
\label{DD2} 2\cdot\frac{d^2\sigma_{SD}}{dtdM_X^2}=
A_{SD}\beta^2(t)\Tilde{W}^{Pp}_2(M^2_X,t)\left(\frac{s}{M_X^2}\right)^{2\alpha_P(t)-2}
\,,
\end{equation}
where $\Tilde{W}^{Pp}_2(M^2_X,t)$ is related to the proton SF, $ F_2^p(M^2_X,t)$ (see Sec.~\ref{sec:proton_sf} for  details).

From Fig.~\ref{Fig:DD}, the differential cross section proton-proton double diffraction (DD) process is: 
\begin{equation}
\label{DD4} \frac{d^3\sigma_{DD}}{dtdM_X^2dM_Y^2}=
A_{DD}\Tilde{W}^{Pp}_2(M^2_X,t)\Tilde{W}^{Pp}_2(M^2_Y,t)\left(\frac{ss_0}{M_X^2 M_Y^2}\right)^{2\alpha_P(t)-2}
,   
\end{equation}
where $\Tilde{W}^{Pp}_2(M^2_X,t)$ is the same function as that used in the $SD$ reaction, with corresponding arguments. 

Accordingly, 
%to Figs.~\ref{fig:diffproc_fact_4}, \ref{fig:diffproc_fact_5} and \ref{fig:diffproc_fact_6}. 
the differential cross sections of proton-proton central diffraction (CD), central diffraction with single diffraction (CDS) and central diffraction with double diffraction (CDD):
\begin{equation}
\label{DD5} \frac{d^4\sigma_{CD}}{dt_1dt_2d\xi_1d\xi_2}=A_{CD}\beta^2(t_1)\beta^2(t_2)\Tilde{W}^{PP}_2(M_Z^2,t_1,t_2)\xi_1^{1-2\alpha_P(t_1)}\xi_2^{1-2\alpha_P(t_2)}\,,
\end{equation}
\begin{align}
\label{DD6} 2\cdot\frac{d^5\sigma_{CDS}}{dt_1dt_2d\xi_1d\xi_2dM_X^2}=A_{CDS}\beta^2(t_2)\Tilde{W}^{Pp}_2(M^2_X,t_1)\Tilde{W}^{PP}_2(M_Z^2,t_1,t_2)\\\nonumber
\times\xi_1^{1-2\alpha_P(t_1)}\left(\frac{s_0}{M^2_X}\right)^{2\alpha_P(t_1)+2}\xi_2^{1-2\alpha_P(t_2)}\,,
\end{align}
\begin{align}
\label{DD7} \frac{d^6\sigma_{CDD}}{dt_1dt_2d\xi_1d\xi_2dM_X^2dM_Y^2}=A_{CDD}\Tilde{W}^{Pp}_2(M^2_X,t_1)\Tilde{W}^{Pp}_2(M^2_Y,t_2)\Tilde{W}^{PP}_2(M_Z^2,t_1,t_2)\\\nonumber
\times\xi_1^{1-2\alpha_P(t_1)}\left(\frac{s_0}{M^2_X}\right)^{2\alpha_P(t_1)+2}\xi_2^{1-2\alpha_P(t_2)}\left(\frac{s_0}{M^2_Y}\right)^{2\alpha_P(t_2)+2}\,,
\end{align}
where\footnote{The equality $M_Z^2=\xi_1\xi_2s$ is not exact. If there are two incoming protons with four-momenta $p_1$ and $p_2$, then $\xi_1p_1$ four-momentum is carried by one of the two Pomerons and $\xi_2p_2$ four-momentum is carried by the other one. Consequently, the squared mass of the centrally produced system is: $M^2_Z=(\xi_1p_1+\xi_2p_2)^2=(\xi_1^2+\xi_2^2)m_p^2+2\xi_1\xi_2(s/2-m_p^2)$, where $m_p$ is the mass of the proton. Using the fact that $m^2_p\ll s$, one has: $M^2_Z\approx\xi_1\xi_2s$.} $M_Z^2=\xi_1\xi_2s$, furthermore $\Tilde{W}^{PP}_2(M^2_X,t)$ is the contribution of the inelastic $PP$ vertex to the differential cross section related to the Pomeron SF, $F^P_2(M_Z^2,t)$ as explained in Sec.~\ref{sec:proton_sf}. Note that $t_1$ and $t_2$ are connected to the virtualities of the colliding Pomerons: $Q_1=-q_1^2=-t_1$ and $Q_2=-q_2^2=-t_2$, where $Q_1$ and $Q_2$ are the virtualities and $q_1$ and $q_2$ are the four momenta of the Pomerons.

\section{The inelastic $Pp$ and $PP$ vertices}\label{sec:proton_sf}

Following Refs.~\cite{Jenkovszky:2010ym, Jenkovszky:2011bt, Jenkovszky:2012hf}, we write the Pomeron-proton vertices as:
\begin{equation}
\Tilde{W_2}^{Pp}(M_X^2,t)\equiv\frac{W^{Pp}_2(M_X^2,t)}{2m_p},
\end{equation}
where
\begin{equation}
W^{Pp}_2(M_X^2,t) = \frac{F^p_2(M_X^2,t)}{\nu(M_X^2,t)},\,\,\,\,\,\,\,\,\,F^p_2(M_X^2,t)={-t(1-x)\over{4\pi \alpha (1-4m_p^2 x^2/t)}}  
\sigma_t^{Pp}(M_X^2,t)~,
\label{eq:DIS_SF}
\end{equation}
$\sigma_t^{Pp}$ is the total Pomeron-proton cross section, $m_p$ is the mass of the proton,  $\alpha$ is the fine structure constant,
\begin{equation}\label{eq:variab_x}
x\equiv x(M_X^2,t)=\frac{-t}{M^2_X-t-m_p^2},
\end{equation}
and
\begin{equation}
\nu(M_X^2,t) = \frac{-t}{2m_p x(M_X^2,t)}.
\end{equation}

The total $Pp$ cross section is:
\begin{equation}
\sigma_t^{Pp}(M_X^2,t)=\sigma_{t,0}^{Pp}(M_X^2)+\sigma_{t,{\rm res}}^{Pp}(M_X^2,t),
\end{equation}
where 
\begin{equation}
\sigma_{t,0}^{Pp}(M_X^2)=\sigma_0 \tau^8(M_X^2) \left( \frac{M_X^2}{s_{0}}\right)^{\epsilon},
\end{equation}
and according to the optical theorem
\begin{equation}
\sigma_{t,{\rm res}}^{Pp}(M_X^2,t)=\frac{8\pi}{P_{CM}M_X}\,{\Im}{\rm m}\  A_{\rm res}^{Pp}(M_X^2,\Tilde{t}=0)~,
\end{equation}
with $\sigma_0= 2.82~\mbox{mb}$ or $7.249~\mbox{GeV}^{-2}$ \cite{Ciesielski:2012mc}, $$\tau(M_X^2)=\frac{e^{-M_X^2/m_0^2}-1}{e^{-M_X^2/m_0^2}+1},\,\,\,\,\,\,m_0^2=1~{\rm GeV^2},$$
$$
P_{CM}\equiv P_{CM}(M_X^2,t)=\frac{M_X^2-m_p^2}{2(1-x)}\sqrt{\frac{1-4m_p^2 x^2/{t}}{M_X^2}},
$$
where $x$ is defined by Eq.(\ref{eq:variab_x}). Here $\tau^8(M_X^2)$ is included \footnote{The power of $8$ is needed to provide sharp enough suppression in the kinematical region where no dissociation occurs.} in $\sigma_{t,0}^{Pp}(M_X^2)$ to suppress it in the region $M_X^2<(m_p+m_{\pi^{0}})^2$ where no dissociation occurs and also in the low $M_X^2$ region where dissociation occurs but resonances do not appear. 

Note that $t\neq\tilde t$. $t$ is connected to the virtuality of the radiated particle, the Pomeron, in the $pp\rightarrow Xp$ process, $Q^2 = - q^2 = -t$, where $q$ is the four-momentum of the Pomeron. $\tilde t$ is the squared four-momentum transfer in the $Pp\rightarrow Pp$ process. Hence, by the optical theorem, $\sigma_{t,res}^{Pp}={\Im}{\rm m}\,  A^{Pp}_{\rm res}(M_X^2,\tilde t=0)$ up to normalization, where ${\Im}{\rm m}\,  A^{Pp}_{\rm res}$ is the imaginary part of the $Pp$ scattering amplitude that includes the resonances. According to  Ref.~\cite{Jenkovszky:2010ym, Jenkovszky:2011bt}, the latter is given as:
\begin{equation}\label{ImA_fin}
{\Im}{\rm m}\,  A^{Pp}_{\rm res}(M_X^2,\tilde t)= \sum_{J}\frac{[f(\tilde t)]^{J+3/2} \Im {\rm m}\alpha_{N^*}(M^2_x)}{(J-\, \Re {\rm e}\alpha_{N^*}(M_x^2))^2+ (\Im {\rm m}\,\alpha_{N^*}(M_x^2))^2},
\end{equation}
where $\alpha_{N^*}$ is the nucleon trajectory, 
\begin{equation}
\label{eq3} f(\tilde t)=(1-\tilde t/t_0)^{-2},
\end{equation}
and $t_0=0.71$ GeV$^2$.  

The explicit form of the nucleon trajectory is given in Refs.~\cite{Jenkovszky:2010ym, Jenkovszky:2012hf}. Resonances on this trajectory appear with total spins  $J = 5/2, 9/2, 13/2$, ... . 

The contribution from the $PP$ vertex to the differential cross section is:
\begin{equation}
\Tilde{W_2}^{PP}(M_Z^2,t_1,t_2)\equiv\frac{F^P_2(M_Z^2,t_1,t_2)}{\nu^P(M_Z^2,t_1,t_2)},
\end{equation}
where
\begin{equation}
F^P_2(M_Z^2,t_1,t_2)=\frac{\nu^P|t_1|}{4\pi^2 \alpha \sqrt{(\nu^P)^2-t_1t_2}}
\sigma_t^{PP}(M_Z^2,t_1,t_2)~,
\label{eq:Pomeron_Pomeron_SF}
\end{equation}
is the Pomeron structure function based on the structure function of the virtual photon given in Ref.~\cite{BERGER19871},

\begin{equation}\label{eq:xP}
\nu^P\equiv \nu^P(M_Z^2,t_1,t_2) = \frac{1}{2}\left(M^2_Z-t_1
  -t_2\right).
\end{equation}

The total $PP$ cross section is:
\begin{equation}
\sigma_t^{PP}(M_Z^2,t_1,t_2)=\sigma_{t,0}^{PP}(M_Z^2)+\sigma_{t,{\rm res}}^{PP}(M_Z^2,t_1,t_2),
\end{equation}
where $\sigma_{t,0}^{PP}$ is identified by $\sigma_{t,0}^{Pp}$ as in Ref.~\cite{Ciesielski:2012mc},
\begin{equation}
\sigma_{t,{\rm res}}^{PP}(M_Z^2,t_1,t_2)=\frac{8\pi}{P_{CM}\sqrt{M_Z^2}}\,{\cal I}m\  A_{\rm res}^{PP}(M_Z^2,\tilde t = 0)~,
\end{equation}
$
P_{CM}\equiv P_{CM}(M_Z^2,t_1,t_2)=\frac{M_Z^2-t_1}{2\left(1+\frac{t_2}{2\nu^P}\right)}\sqrt{\frac{1+t_1t_2/{(\nu^P)^2}}{M_Z^2}}
$
and $\nu^P$ is given by Eq.~(\ref{eq:xP}). Based on Ref.~\cite{Fiore:2017xnx}:
\begin{equation} \label{ImA}
\Im {\rm m}\, A^{PP}_{\rm res}(M_{Z}^2,\tilde t)=\sum_{i=f,P}\sum_{J}\frac{[f_{i}(\tilde t)]^{J+2} 
\Im {\rm m}\,\alpha_{i}(M_{Z}^2)}{(J-\Re{\rm e}\,\alpha_{i}(M_{Z}^2))^2+ 
(\Im {\rm m}\,\alpha_{i}(M_{Z}^2))^2},
\end{equation}
where the index $i$ runs over the trajectories, which contributes to the amplitude. For all trajectories we sum over the states with full spins  $J$. The $f_i(\tilde t)$ is the pole residue and given by Eq.~(\ref{eq3}) for all trajectories uniformly. Note that $\tilde{t}$ is the squared four-momentum transfer in the $PP\rightarrow PP$ process while $t_1$ and $t_2$ are connected to the virtualities of the colliding Pomerons.

The $PP\rightarrow M_Z^2$ Pomeron-Pomeron channel couples to the Pomeron and the $f$-meson by the conservation of the quantum numbers. The explicit form of the Pomeron trajectory can be found in Ref.~\cite{Szanyi:2019kkn}, while that of the $f$-meson trajectories are given in Ref.~\cite{Fiore:2017xnx}. At the present stage of research we include only glueballs lying on the Pomeron trajectory. Ordinary mesons will be added in a forthcoming study.

\section{Integrated cross sections}\label{sec:intcs}

In this section integrated cross sections for the $SD$, $DD$ and $CD$ reactions are presented. Numerical calculations for $CDS$ and $CDD$ processes are postponed to a later study.

\begin{figure}[hbt!]
\begin{center}
\includegraphics[clip,scale=0.4]{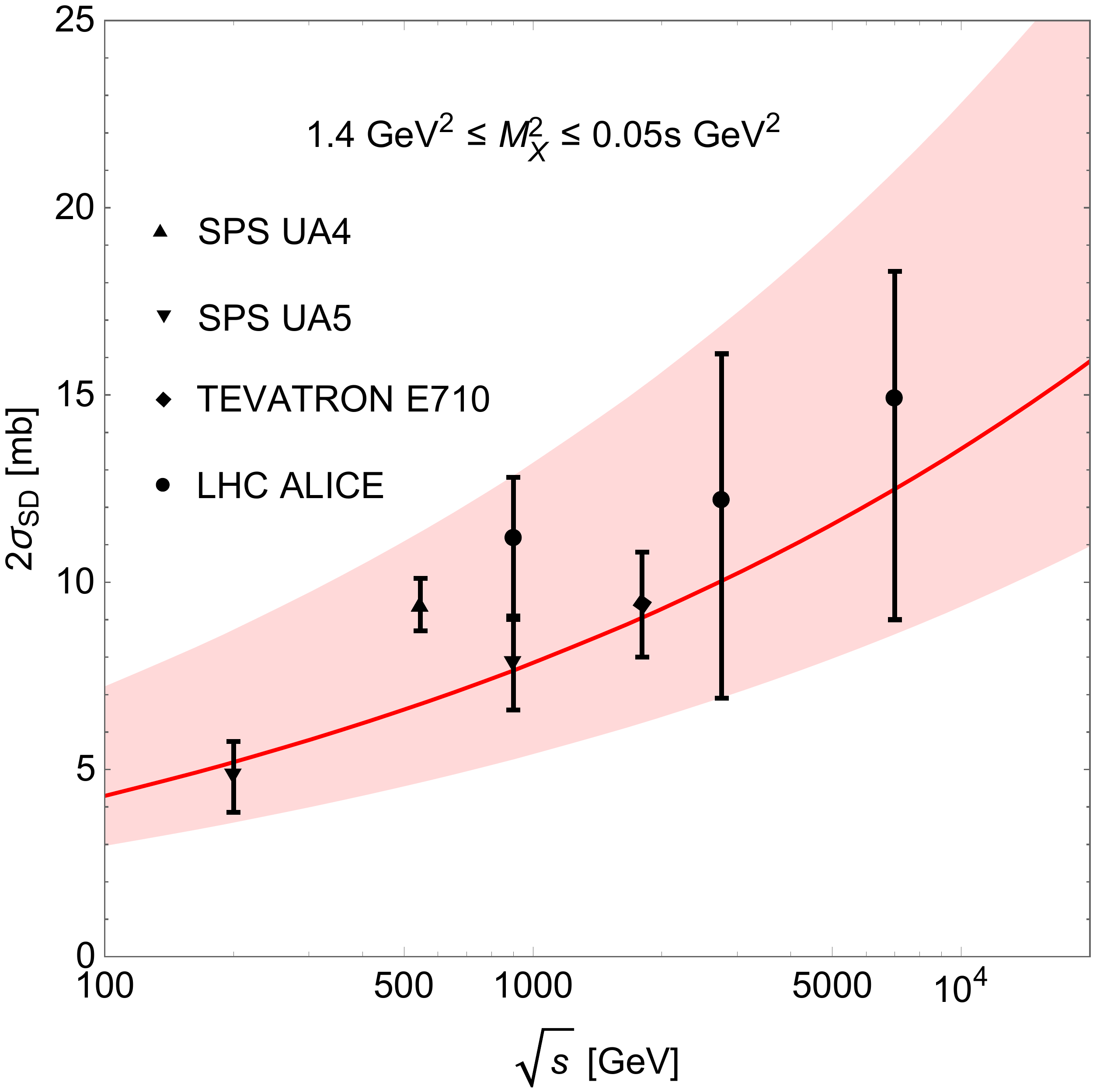}
\end{center}
%\vspace{-10cm}
\caption{Integrated SD cross section. The shaded area corresponds to the uncertainty arising from the normalisation parameter $A_{SD}$.}
\label{Fig:Integrated_SD}
\end{figure}

For $SD$ we have
\begin{equation}
2\sigma_{SD} = \int_{{M^2_X}_{min}}^{{M^2_X}_{max}} dM_X^2 \int_{t_{min}}^{t_{max}} dt ~2 \cdot\frac{d\sigma_{SD}^2}{dM_X^2dt},
\end{equation}
where ${M^2_X}_{min} = 1.4$ GeV$^2$ \cite{Goulianos:1995vn}, ${M^2_X}_{max} = 0.05s$ GeV$^2$, $t_{min}=-\infty$ and $t_{max}=0$ GeV$^2$ (practically $t_{min}=-1$ GeV$^2$). The result is shown in Fig.~\ref{Fig:Integrated_SD} with $A_{SD}=0.063^{+0.043}_{-0.020}$ mb/GeV$^2$ resulting from a fit to the experimental data. The theoretical uncertainties in Fig. 2 are correlated with
the errors in the data.

\begin{figure}[h]
\begin{center}
\includegraphics[clip,scale=0.4]{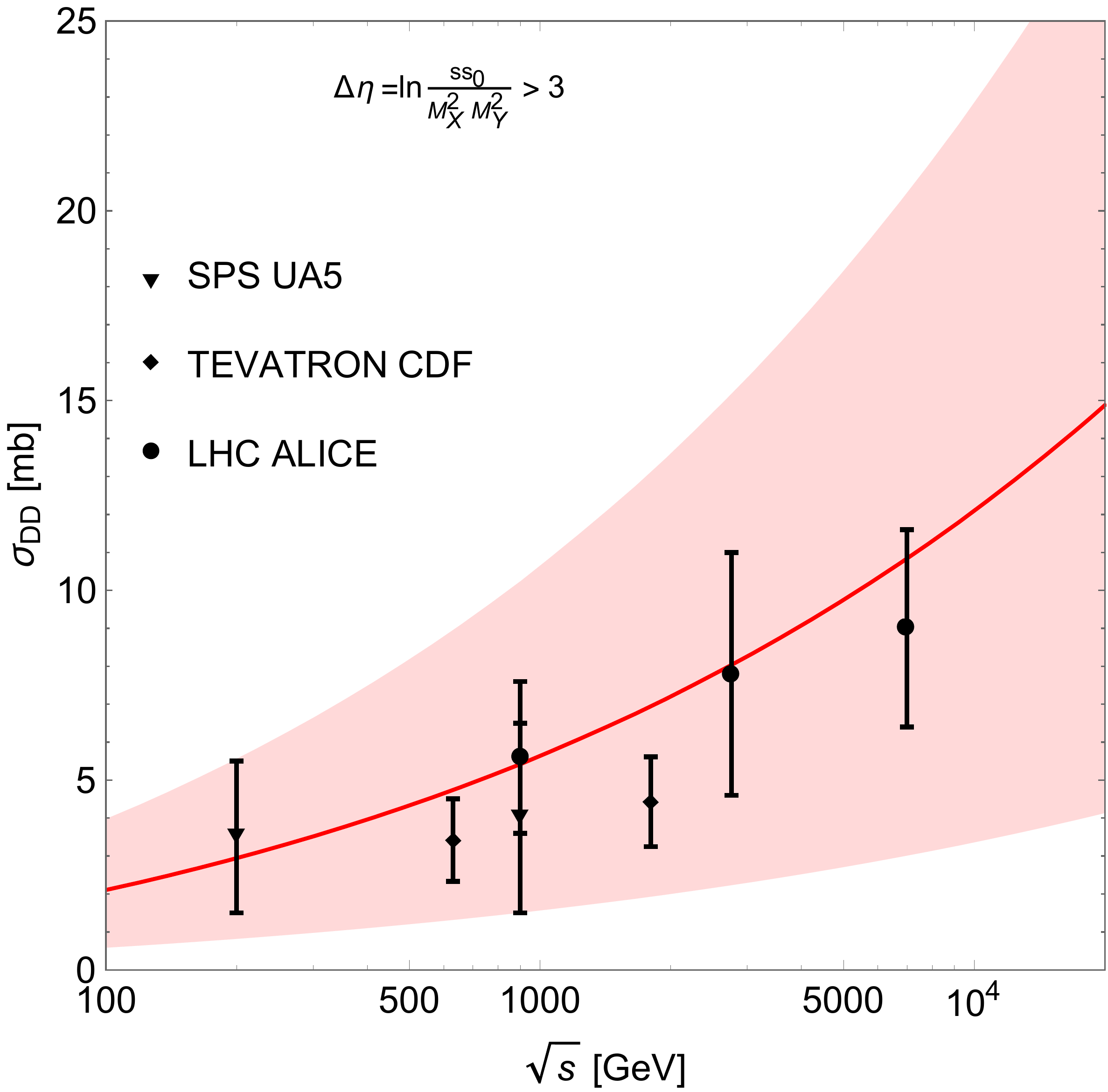}
\end{center}
%\vspace{-10cm}
\caption{Integrated DD cross section. The shaded area corresponds to the uncertainty arising from the normalisation parameter $A_{DD}$.}
\label{fig:DD_tot}
\end{figure}

For $DD$  one has \cite{Goulianos:1995vn}:
\begin{equation}
\sigma_{DD} = \int_{{M^2_X}_{min}}^{{M^2_X}_{max}} dM_X^2 \int_{{M^2_Y}_{min}}^{{M^2_Y}_{max}} dM_Y^2 \int_{t_{min}}^{t_{max}} dt \frac{d\sigma_{DD}^3}{dM_X^2dM_Y^2dt},
\end{equation}
where ${M^2_X}_{min} = 1.4$ GeV$^2$, ${M^2_X}_{max}= 0.05ss_0/{M^2_Y}_{min}$  GeV$^2$, ${M^2_Y}_{min}=1.4$ GeV$^2$, ${M^2_Y}_{max}= 0.05ss_0/{M_Y}^2_{min}$ GeV$^2$, $s_0=1$ GeV$^2$, $t_{max}=0$ GeV$^2$ and $t_{min}=-\infty$ . The result is shown in Fig.~\ref{fig:DD_tot} with $A_{DD}=9^{+8.0}_{-6.5}\times 10^{-5}$ mb/GeV$^2$ resulting from a fit to the experimental data.

For CD it is convenient to use the variables $\Delta \eta = \ln\frac{s}{M^2_Z}$ (rapidity gap) and $\eta_c$ (the center of the
centrally-produced system in $\eta$) \cite{Goulianos:2004as, Goulianos:1995vn}:
\begin{align}
\label{DD5m} \frac{d^4\sigma_{CD}}{dt_1dt_2d\Delta\eta d\eta_c}=A_{CD}\beta^2(t_1)\beta^2(t_2)\Tilde{W}^{PP}_2\left(se^{-\Delta\eta},t_1,t_2\right)\\\nonumber \times e^{\frac{1}{2}\left[\alpha_P(t_1)-1\right]\left[\Delta\eta+\eta_c\right]}e^{\frac{1}{2}\left[\alpha_P(t_2)-1\right]\left[\Delta\eta-\eta_c\right]}\,.
\end{align}
Now the integrated cross section for CD is
\begin{equation}
\sigma_{CD} =  \int_{{t_1}_{min}}^{{t_1}_{max}} dt_1 \int_{{t_2}_{min}}^{{t_2}_{max}} dt_2 \int_{\Delta\eta_{min}}^{\Delta\eta_{max}} \int_{{\eta_c}_{min}}^{{\eta_c}_{max}} \frac{d^4\sigma_{CD}}{dt_1dt_2d\Delta\eta d\eta_c},
\end{equation}
where ${t_1}_{min} = {t_2}_{min} = -\infty$, ${t_1}_{max} = {t_2}_{max} = 0$ GeV$^2$, $\Delta\eta_{min}=3$, $\Delta\eta_{max}=\ln(s/s_0)$, $s_0=1$ GeV$^2$, ${\eta_c}_{min}=-\frac{1}{2}\left(\Delta\eta-\Delta\eta_{min}\right)$ and ${\eta_c}_{max}=\frac{1}{2}\left(\Delta\eta-\Delta\eta_{min}\right)$ \cite{Ciesielski:2012mc}. 

\begin{figure}[hbt!]
\begin{center}
\includegraphics[clip,scale=0.48]{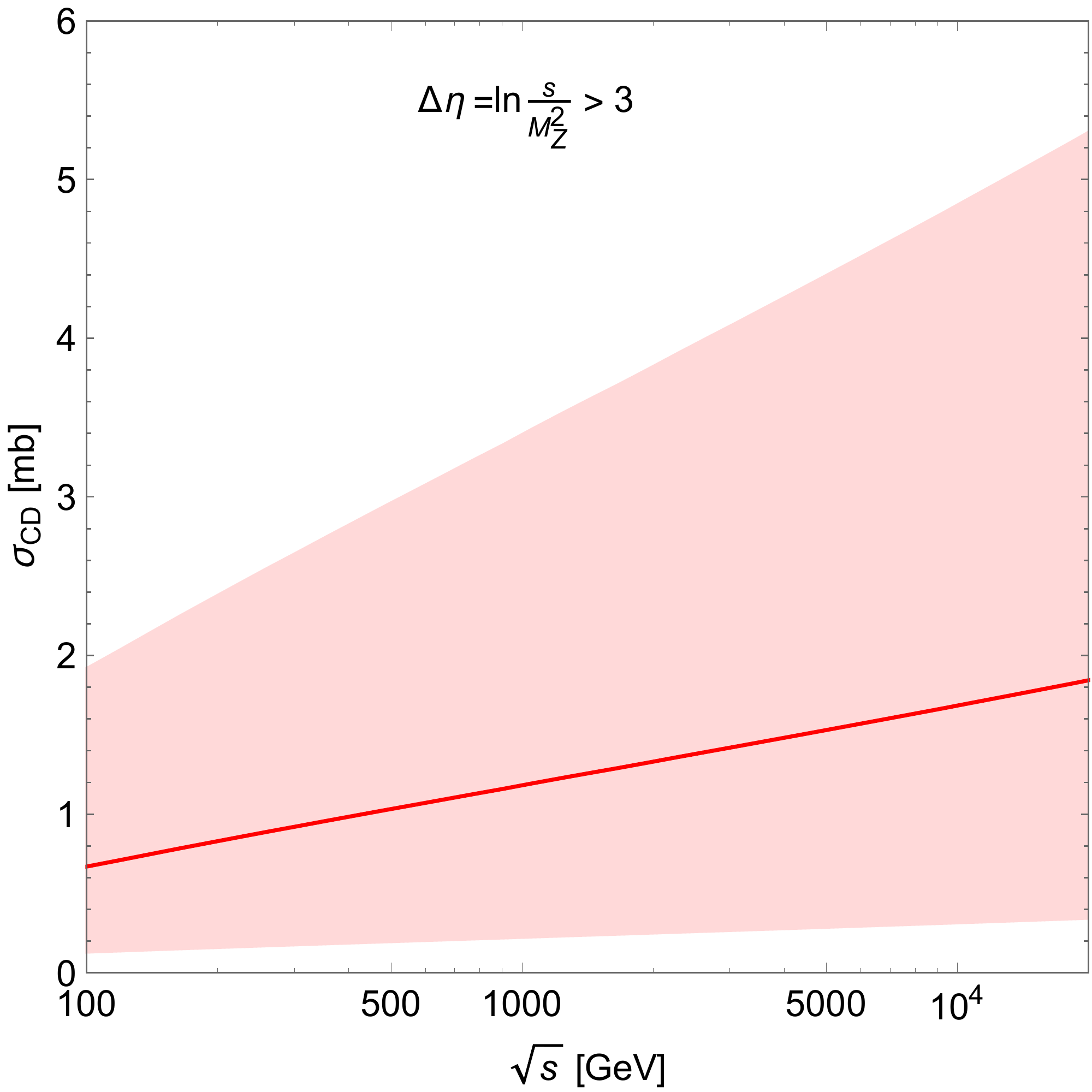}
\end{center}
	\caption{Integrated CD cross section. The shaded area corresponds to uncertainties inherent in the normalisation parameter $A_{CD}$.\label{fig:CD_tot}}
\end{figure}

The results are shown in Fig.~\ref{fig:CD_tot} with $A_{CD}=0.066^{+0.124}_{-0.54}$ mb/GeV$^2$. The value of this normalisation parameter is obtained using the relation $\sigma_{CD}\approx \frac{(2\sigma_{SD})^2}{\sigma^{pp}_{tot}}$ based on Regge factorisation. The uncertainty is obtained by the calculated uncertainty of $2\sigma_{SD}$ and the total experimental uncertainty of $\sigma^{pp}_{tot}$ \cite{TOTEM:2013vij} at 7 TeV. 

%The result is shown in Fig.~ 
%\ref{fig:CDtot} without the $f$-%meson contribution and with 
%$A_{CD}=1.28\times 10^{-2}$ $
%(given that $\sigma_{CD}\approx%\sigma_{SD}^2/\sigma_{SD}^{pp}$).

\section{Predictions for differential cross sections}\label{sec:diffcs}

This section is devoted to our predictions for $SD$, $DD$ and $CD$ multiple differential cross sections at $\sqrt{s}=14$ TeV in the low-mass region. 

\begin{figure}[hbt!]
\centering
\includegraphics[clip,scale=0.42]{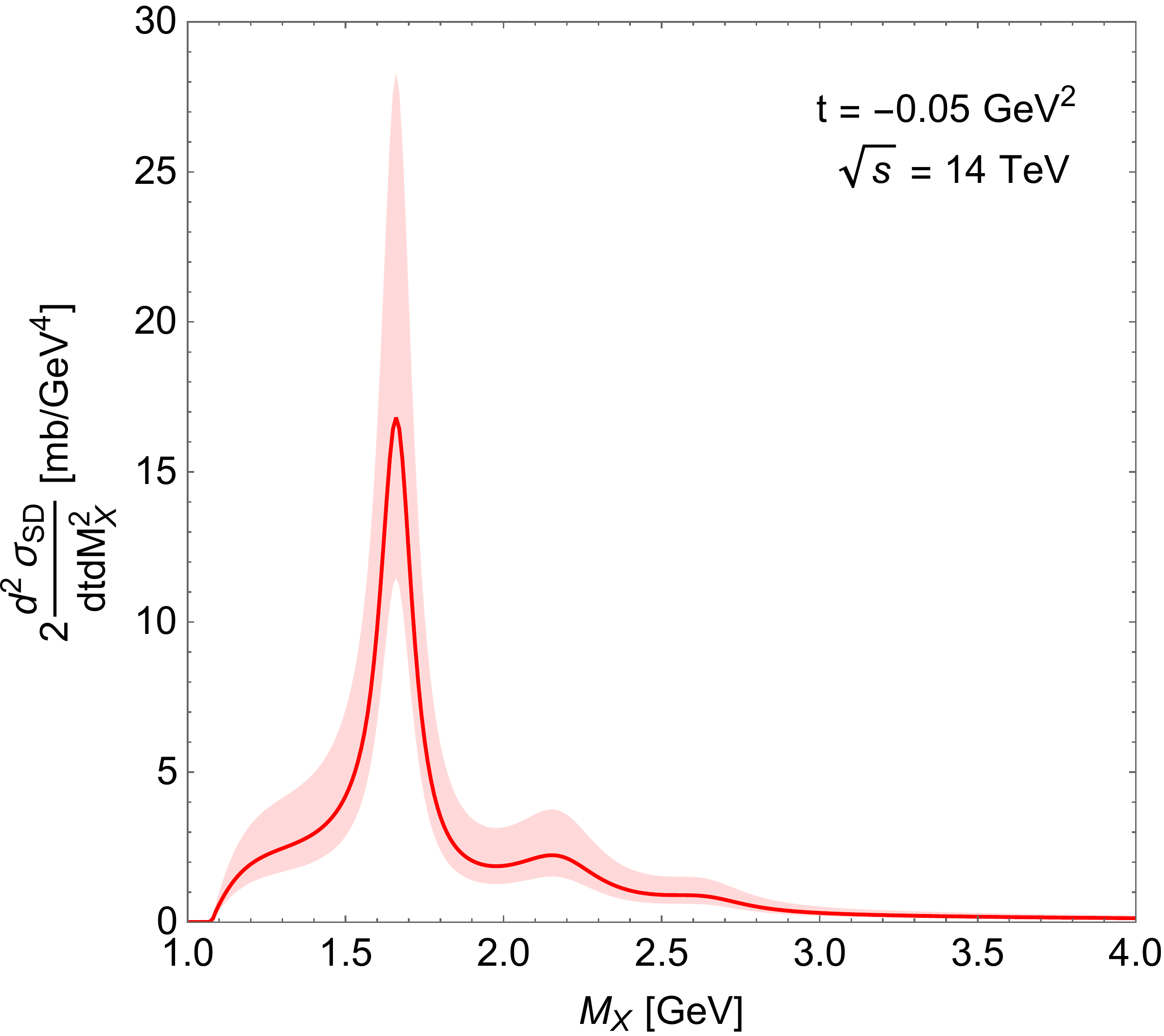}
%\vspace{-10cm}
\caption{Mass dependence of  the $SD$ double differential cross section.}
\label{Fig:SD_M}
\end{figure}

The $M_X$ dependence of $SD$ double differential cross section is shown in Fig. \ref{Fig:SD_M}. The visible peaks correspond to nucleon resonances: $N^*(1680)$, $N^*(2220)$, and $N^*(2700)$. Fig. \ref{Fig:SD_M} shows the squared momentum transfer dependence of this cross section: a peak at low-$|t|$ followed by the usual exponential decrease. The shaded areas around the curve shows the uncertainty of the calculations following from the uncertainty of the normalisation parameter.

 \begin{figure}[hbt!]
\centering
\includegraphics[clip,scale=0.42]{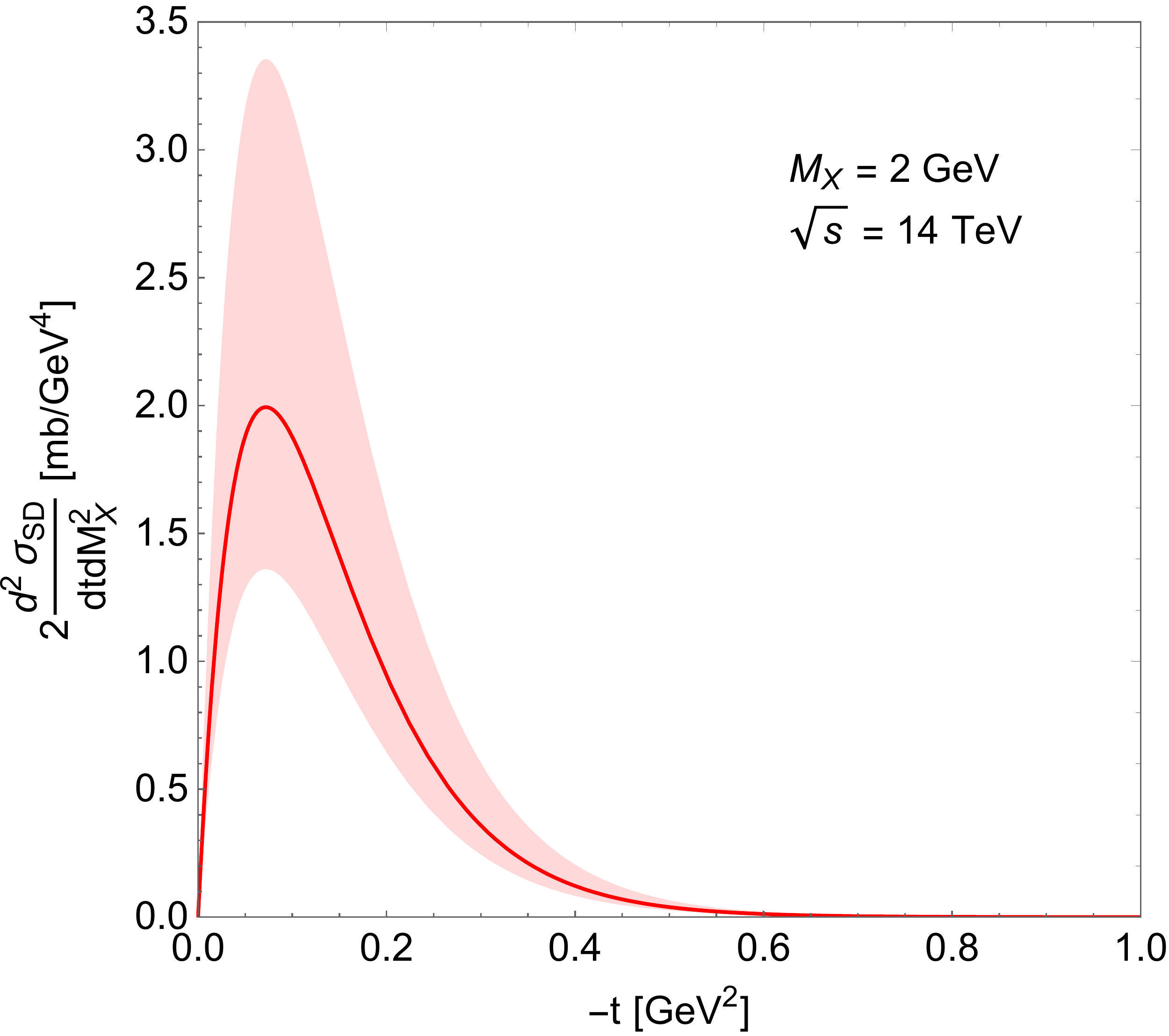}
%\vspace{-10cm}
\caption{Squared momentum transfer dependence of the $SD$ double differential cross section.}
\label{Fig:SD_t}
\end{figure}

\begin{figure}[hbt!]
\begin{center}
\includegraphics[clip,scale=0.55]{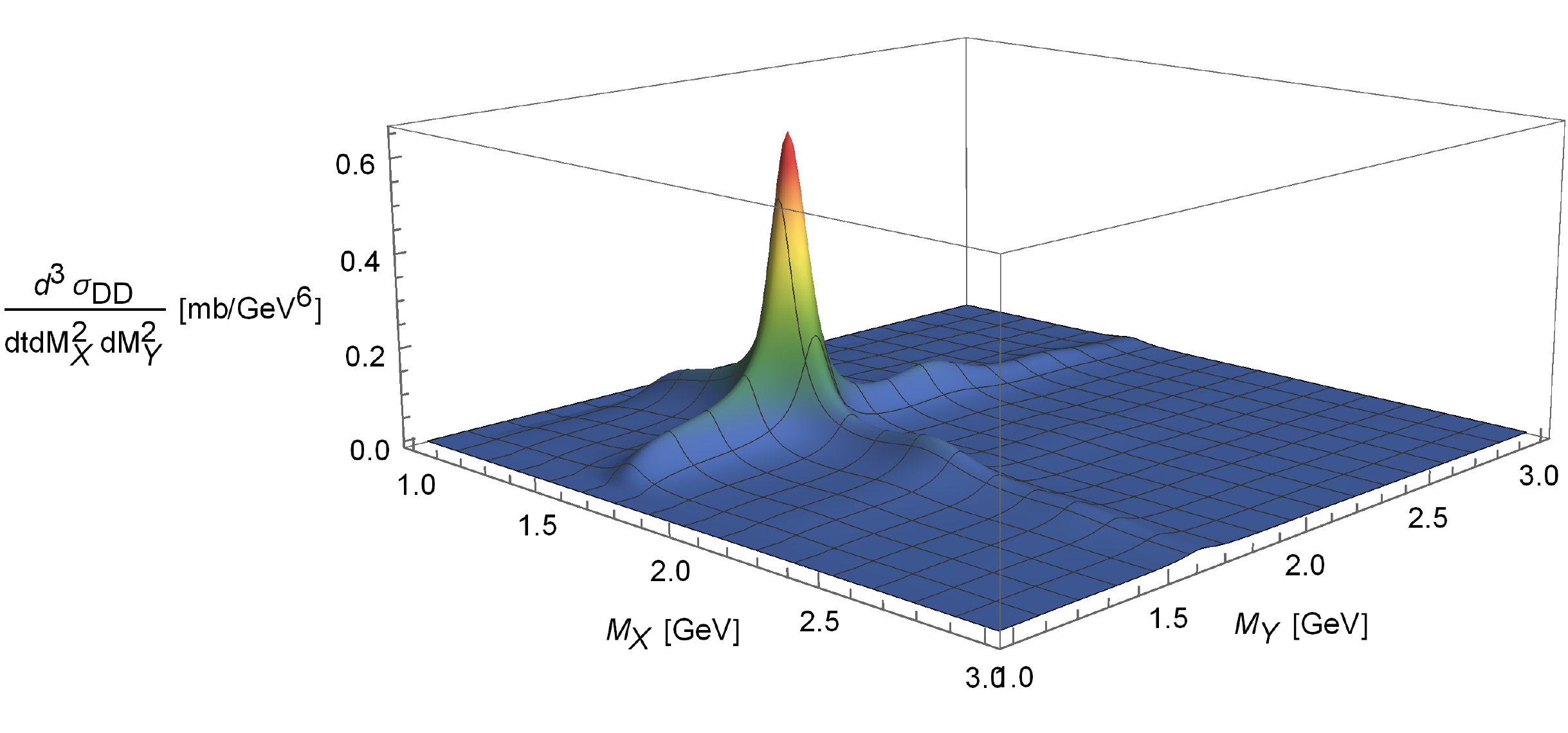}
\includegraphics[clip,scale=0.5]{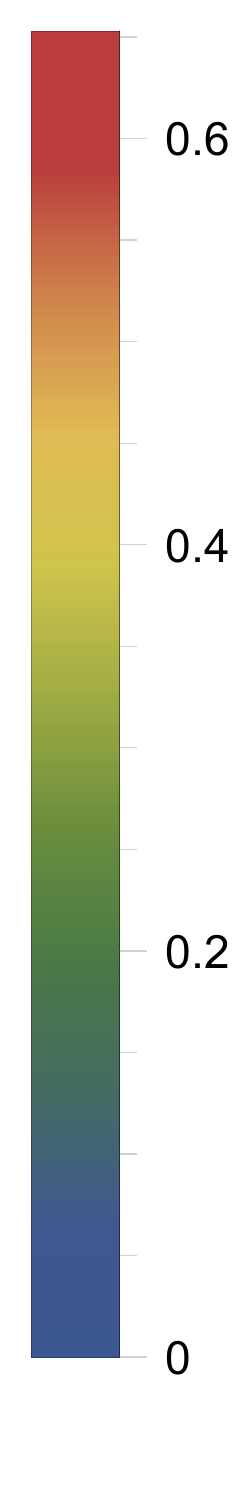}
\end{center}
%\vspace{-10cm}
\caption{Mass dependence of the $DD$ triple differential cross section.}
\label{Fig:DD1}
\end{figure}

\begin{figure}[hbt!]
\begin{center}
\includegraphics[clip,scale=0.42]{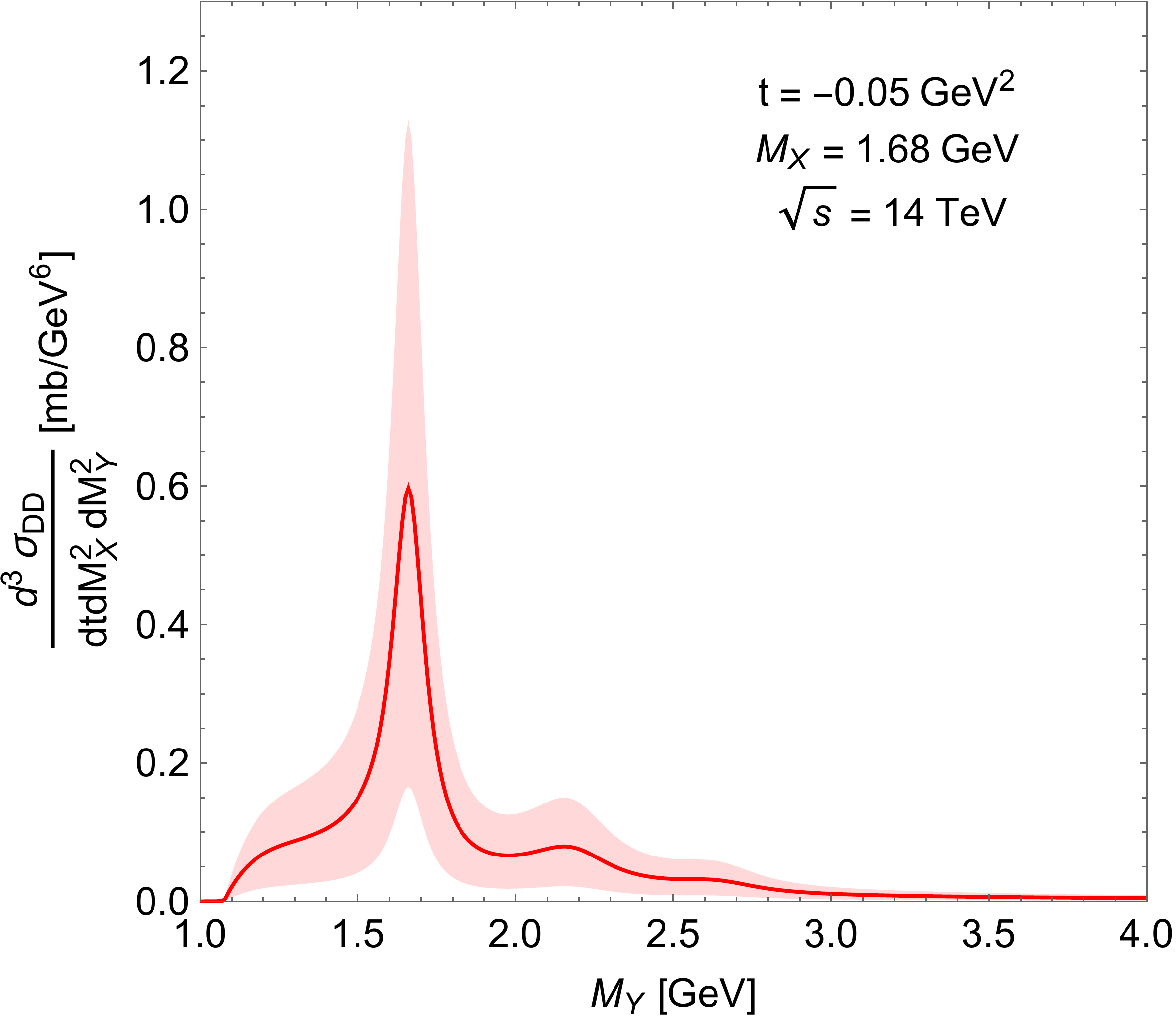}
\end{center}
%\vspace{-10cm}
\caption{Same as Fig.~\ref{Fig:DD1} calculated at $M_X=1.68$ GeV.}
\label{Fig:DD2}
\end{figure}

The $M_X$ and $M_Y$ dependence of the $DD$ triple differential cross section is shown in Fig. \ref{Fig:DD1} as a surface. Similar to $SD$, the peaks correspond to nucleon resonances: $N^*(1680)$, $N^*(2220)$, and $N^*(2700)$. Fig. \ref{Fig:DD2} is a "slice" of Fig. \ref{Fig:DD1} corresponding to a fixed $M_X$ showing the uncertainty of the calculation originating from the uncertainty of the normalisation parameter.

\begin{figure}[hbt!]
	\centering
	\includegraphics[clip,scale=0.52]{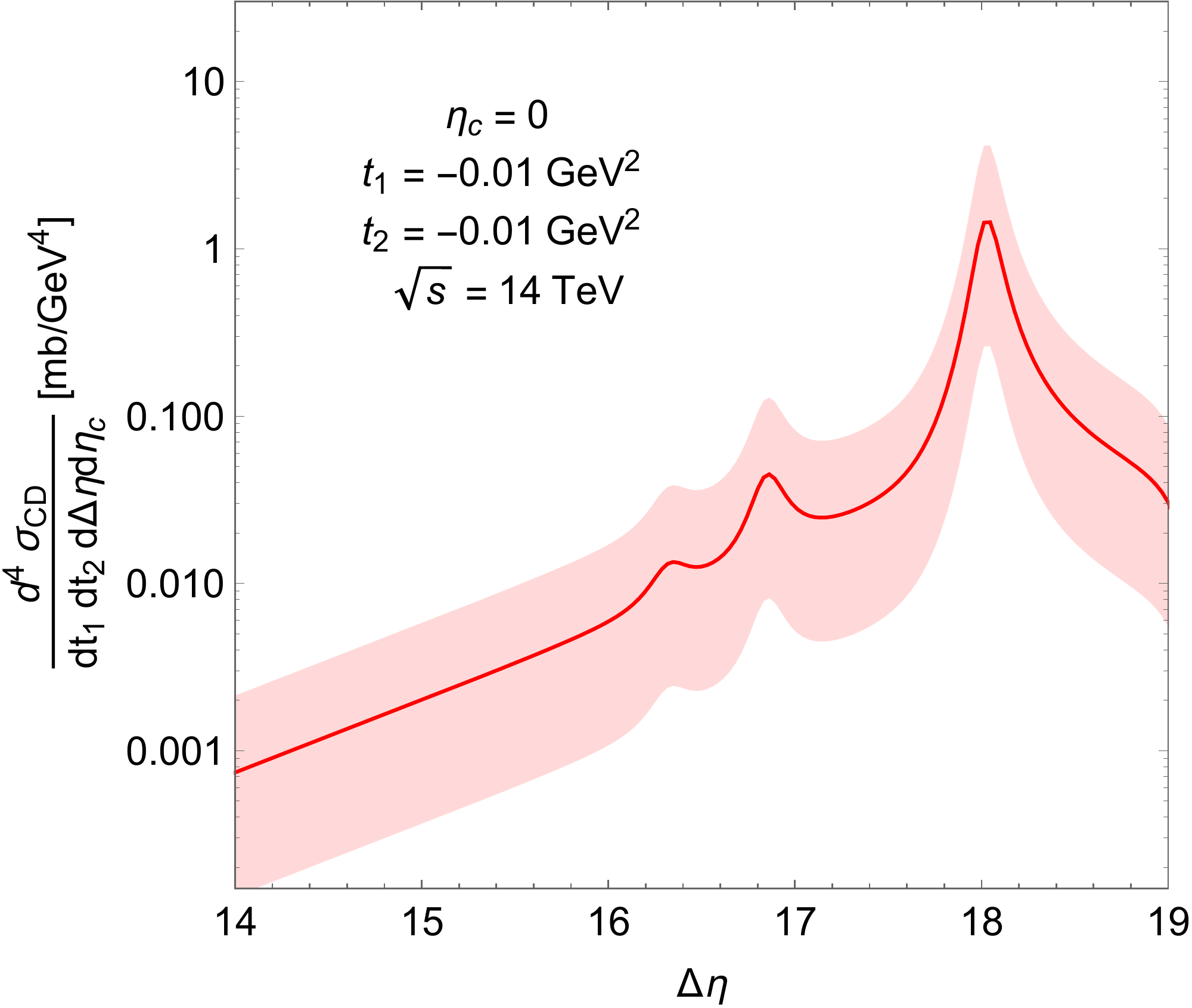}
	\caption{Mass dependence of the $CD$ quadruple differential cross section.\label{fig:CD_diff}}
\end{figure}

The $M_Z$ dependence of the $CD$ quadruple differential cross section is shown in Fig.~\ref{fig:CD_diff}. The visible peaks correspond to glueball resonances lying on the Pomeron trajectory: $J^{PC}=2^{++},~4^{++}$, and $6^{++}$. Mesons will be included in a forthcomng study.

\newpage
\section{Summary}\label{sec:concl}

In this paper we presented updated results on modelling single and double diffraction as well as novel results on modelling central diffraction. The modelling is based on Regge factorisation accompanied by the identification of the contributions of inelastic vertices by structure functions.

%In the present paper we have summarized the present status of the important and rapidly developing subject of nucleon diffractive dissociation ot high energies, typical of the LHC.

We stress that one of the main unknown objects is the inelastic $Pp$ vertex. As mentioned in the Introduction, in most of the papers on the subject, e.g. in Refs. \cite{Goulianos:1982vk, Goulianos:2001hck, Goulianos:2004as, Goulianos:1995vn, Goulianos:2018atd} one associates (following the ideas of Ref.~\cite{Ingelman:1984ns}) the Pomeron with a flux radiated by the incoming proton. The authors of Refs. \cite{Jenkovszky:2010ym, Jenkovszky:2011bt, Jenkovszky:2012hf}, following \cite{PhysRevD.10.170}, take a different viewpoint and identify the  inelastic $Pp$ vertex with the proton SF, known from deep-inelastic electron-proton scattering \cite{Fiore:2003dg}. In Refs.~\cite{Jenkovszky:2010ym, Jenkovszky:2011bt, Jenkovszky:2012hf, Jenkovszky:2013xny} this SF is specified by the direct-channel resonance diagrams dominated by relevant baryon trajectories producing excited nucleon states (mainly $N^*$ resonances).

A completely novel result of this paper is the identification of the inelastic $PP$ vertex with a Pomeron SF. The Pomeron SF is constructed based on the virtual photon SF \cite{BERGER19871} in a way it can contain mesonic and glueball resonances. The treatment of the inelastic $PP$ vertex is crucial in central diffractive dissociation (diagrams 4-6 in Fig.\ref{Fig:DD}). They contain a subdiagram corresponding to collision of two Pomerons (or, more generally, reggeons). Construction of amplitudes describing scatting of virtual hadrons (by "virtual hadrons" we mean  states lying on the Pomeron (or any reggeon) trajectory) is of course an open problem. Our present approach is one possibility although experimental data on central diffraction is needed for justification or for further guide in theoretical developments.

Finally, we highlight that the main part of the dynamics in diffractive dissociation is carried by the Regge trajectories, \textit{i.e.}, nonlinear complex functions. The construction of explicit models of such trajectories is a basic part of this approach, deserving further studies.

\section*{Acknowledgents}
The present paper is based on the talks at the the Zim\'anyi Winter School in Budapest, December 2021 and a Workshop on Diffraction and Femtoscopy in Gy\"ongy\"os, Hungary, May 2022. L.J. thanks Professor Tam\'as Cs\"org\H{o} for inviting him to these interesting events as well as the Organizers for the financial support.  The work was supported by the NKFIH Grant no. K133046. I.Sz. was supported by the M\'arton \'Aron Szakkoll\'egium program. L.J. was supported by the Ukrainian Nat. Ac. Sc. program "Fundamental properties of matter", Grant 1230/22-1. We  thank  O. Skorenok for his collaboration at an earlier stage of this study.

\bibliographystyle{ws-ijmpa}
\bibliography{mybibfile}
%\printbibliography
\end{document}